\documentclass[12pt]{article}

\usepackage{epsfig}
\usepackage{a4}

\newcommand{\half}{\mbox{${\textstyle \frac{1}{2}}$}}
\newcommand{\fmn}[2]{\mbox{${\textstyle \frac{#1}{#2}}$}}
\newcommand{\bbox}[1]{\mbox{\boldmath${#1}$}}

\begin{document}
\setlength{\unitlength}{1mm}
\mbox{ }
\vspace{20mm}

\includegraphics{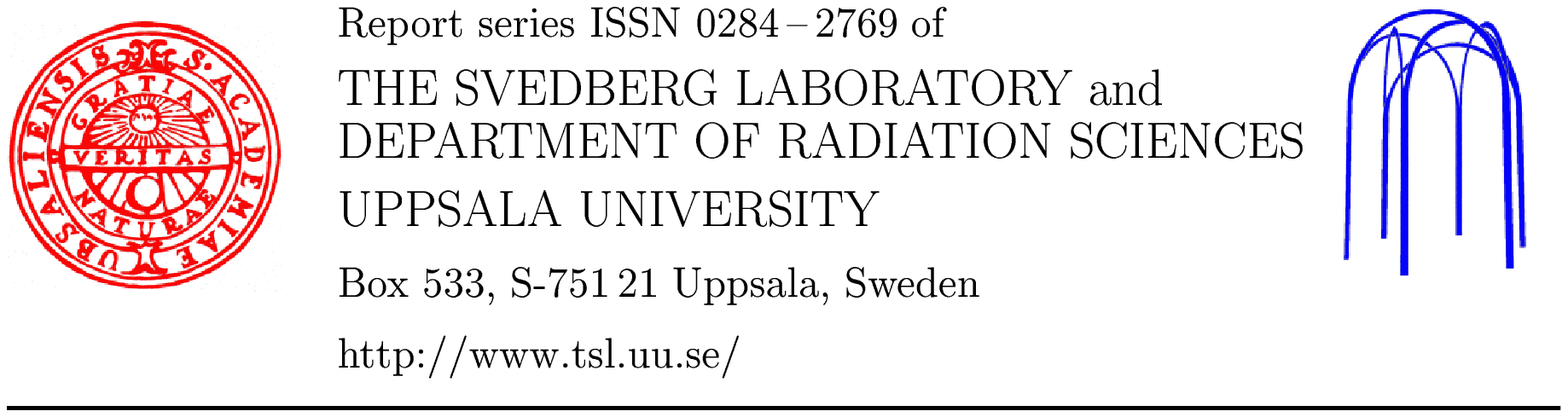}

\begin{flushright}
\begin{minipage}[t]{37mm}
{\bf TSL/ISV-98-0201 \\
December 1998}
\end{minipage}
\end{flushright}

\vspace{10mm}

\begin{center}
{\LARGE\bf Structure in two-pion production in the} \\[2ex]
{\LARGE\bf $\bbox{dd\rightarrow\alpha{X}}$ reaction}
\vspace{20mm}

{\Large Anders G{\aa}rdestig\footnote{grdstg@tsl.uu.se} and
G\"oran F\"aldt\footnote{faldt@tsl.uu.se}}\\[1ex]
Division of Nuclear Physics, Uppsala University, Box 535,\\
S-751 21 Uppsala, Sweden\\[4ex]
{\Large Colin Wilkin\footnote{cw@hep.ucl.ac.uk}}\\[1ex]
University College London, London WC1E 6BT, UK\\
\end{center}

\vspace{10mm}

{\bf Abstract:}
A model for the $dd\rightarrow\alpha\pi\pi$ reaction, based on two parallel
$N\!N\rightarrow{d}\pi$ reactions, is extended to incorporate a complete set of
input amplitudes. While low-energy cross sections are underestimated, the rich
structure observed in the $\alpha$-particle momentum distributions for
$0.8<T_{d}<1.9$~GeV (the ABC effect) is extraordinarily well reproduced. In
addition, a recent measurement of deuteron analyzing powers agrees quite well
with our predictions, both in frequency and magnitude of the oscillations.
\vfill
\clearpage
\setcounter{page}{1}

\section{Introduction}
\label{sec:intro}

On many occasions, interesting physics has been obtained from detailed studies 
of momentum distributions from nuclear reactions. Excitement once arose from the
unexpected width of $\beta$-decay spectra, which led Pauli to postulate the 
existence of the neutrino. In most cases, however, the interest has focused on 
details of the distributions, such as an enhancement or a suppression at a 
certain missing mass, or skewness, or other irregularities. These deviations 
from phase space mark the discoveries of new particles and resonances, increased
propensity for a particular partial wave, or some other peculiarity. Regardless
of the problem at hand, further investigations often proved fruitful and well 
worth the effort. A particular interesting problem of this kind is the striking 
phenomenon observed in several experiments on the
$pn\rightarrow dX$~\cite{HH,Ban71,Plouin}, 
$pd\rightarrow\,^{3}{\rm He}X$~\cite{ABC,Ban71,Ban73}, and
$dd\rightarrow\alpha X$~\cite{Ban76,SPESIII} reactions. In all three cases there
is a peak just above the two-pion threshold with a missing mass 
$M_{X}\approx$300--320~${\rm MeV/c}^2$. Since its position and width vary with 
kinematical conditions, it could not be a new particle or resonance. The only 
alternative is some sort of kinematical enhancement in double pion production. 
The experiments reveal that the effect only occurs in the isospin $I_{X}=0$ 
channels of the two first reactions; the $dd\rightarrow\alpha X$ reaction leads
to a pure $I_X=0$ state, which is probably the reason for the more pronounced
peaks in this case. In association with the peaks at low $\pi\pi$ mass there are
also broad bumps at maximal missing masses. These are most clearly seen in the 
$pn\rightarrow dX$ and $dd\rightarrow\alpha X$ reactions, which could be due to
more symmetric kinematics. These characteristics are collectively known as the 
ABC effect after the discoverers~\cite{ABC}. The suggestion of the original 
authors that the peaks are due to a large $\pi\pi$ $s$-wave scattering length 
was soon ruled out by separate experiments~\cite{Maung} and theoretical 
calculations~\cite{Weinberg}. In addition the central bump (unknown at the time)
cannot be explained in this way, since it would mean that parallel and 
anti-parallel pions are simultaneously favored by the $\pi\pi$ interaction, 
which is very unlikely. The mass of the central bump is also changing 
monotonically with beam energy. It is thus impossible to explain the ABC 
enhancement in terms of some peculiarity of the $\pi\pi$ interaction and one has
to include effects resulting from the presence of other particles in the
reactions and consider the structure of the single pion production mechanism 
itself. In all three reactions there are indications that the ABC effect 
disappears at low beam energies~\cite{Hollas,MOMO,Chap,Barg}, suggesting that 
the subprocesses in any explanation should be strongly energy and angular 
dependent in order to reproduce the quite different dynamics observed in the 
different energy regions.

For the $pn\rightarrow dX$ reaction, on which theoretical investigations have 
mainly focused, the most promising approach is the 
$\Delta\Delta$-model~\cite{BRS}. Here two independent pion productions are
achieved through the excitation of both incident nucleons into $\Delta$-isobars
via meson exchange, with the deuteron then being formed through a final state
interaction of the recoil nucleons arising from the $\Delta$ decay. By using the
energy and angular dependences of $\Delta$-excitation and decay, the authors
could reproduce the shape quite well,  but underestimate the cross section at
larger angles~\cite{Plouin}. A double-nucleon exchange model, also proposed in
the seventies~\cite{ALS} and using separate diagrams for the three different
peaks, showed a much poorer angular distribution~\cite{Plouin} and the authors
were incapable of obtaining an absolute normalization. In his survey of all ABC
experiments and models, Barry~\cite{Barry} tried to improve the original
$\Delta\Delta$-model and to extend it to $pd\rightarrow\,^{3}{\rm He}X$ but 
with little success; his model has, in particular, problems in reproducing the
spectra away from the ABC peaks. Better results were achieved in a chiral bag
model~\cite{KE}, taking into account 80 diagrams to fourth order in  the
$\pi{N\!N}$ coupling constant, but this model still failed in the details. A 
model based on excitation of the $N^{\ast}(1440)$ (Roper) resonance has recently
been proposed for the lower energies~\cite{Luis}. The calculations are in good
agreement with data~\cite{Hollas}, but the contribution from $\Delta\Delta$ 
excitations is negligible. The situation at low energies is thus improving and
might form a basis for a renewed investigation of the ABC effect in the
$pn\rightarrow dX$ reaction.

Despite these efforts, there is as yet no completely satisfactory 
implementation of a model for the ABC enhancement in either the 
$pn\rightarrow dX$ or $pd\rightarrow\,^{3}{\rm He}X$\ reactions. On the other 
hand, the $dd\rightarrow\alpha X$ reaction has not been subjected to any
theoretical investigation because of its perceived complexity. In a preliminary
study~\cite{letter} we were able to reproduce very well the angular dependence
measured at $T_{d}=$~1.25 GeV~\cite{Ban76} using a version of a $\Delta\Delta$
model where independent single pion productions arise from two 
$N\!N\rightarrow d\pi$ reactions occurring  in parallel. In this work, only the 
dominant $N\!N\rightarrow d\pi$ input amplitude was kept and we now lift this 
restriction by including a complete set of amplitudes. This makes it possible to
reproduce all the major features observed in the $dd\rightarrow\alpha X$ 
reaction throughout the $\Delta$ region, including also the recent measurement 
of deuteron analyzing powers~\cite{SPESIII}. The main idea behind the model is 
explained in Sec.~\ref{sec:model}. The general description is then given a 
precise formulation in Sec.~\ref{sec:formal}, to be followed by some remarks 
about the numerical calculations in Sec.~\ref{sec:numerics}. Our predictions are
compared with data in Sec.~\ref{sec:results} and conclusions and outlook are 
given in Sec.~\ref{sec:concl}. Some formulae for the $N\!N\rightarrow d\pi$ 
reaction, in particular the relations between the partial wave and spin 
amplitudes, are collected in an appendix.

\section{Model for $\bbox{\lowercase{dd}\rightarrow\alpha\pi\pi}$}
\label{sec:model}

\begin{figure}[p]
\centering \includegraphics*[39mm,126mm][159mm,186mm]{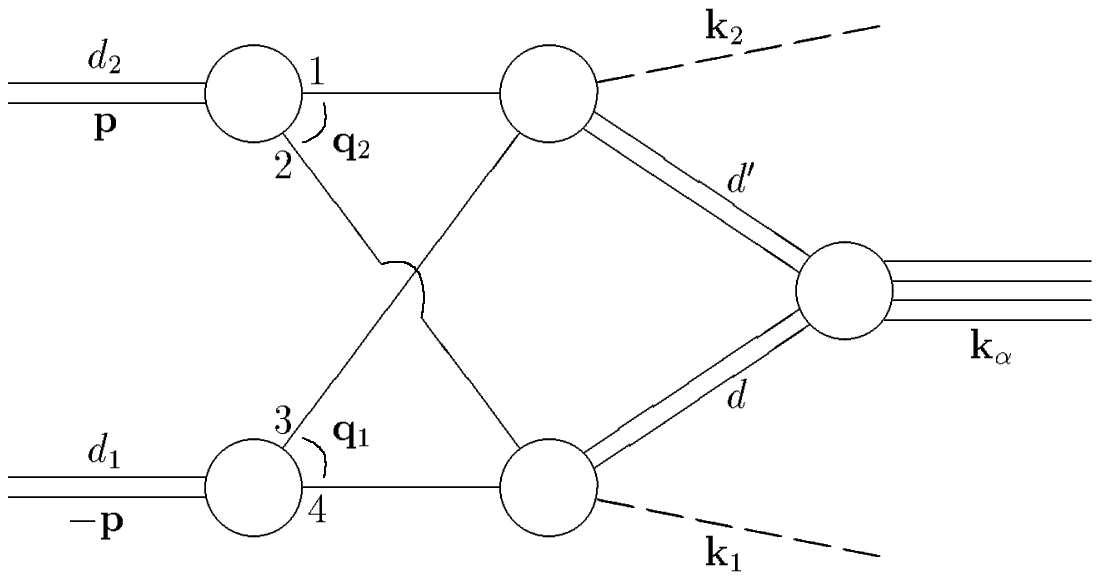}
\caption{Feynman diagram for the $dd\rightarrow\alpha X$ reaction showing the
momenta in the overall cm system}
\label{fig:ddapp}
\end{figure}

Our model for the $dd\rightarrow\alpha\pi\pi$ reaction is based on a simple, 
semi-classical picture and illustrated as a Feynman diagram in 
Fig.~\ref{fig:ddapp}. The two-pion production in the $dd$ collision is viewed as
two free, parallel, and independent $N\!N\rightarrow d\pi$ reactions taking 
place between separate pairs of nucleons from the two deuterons. The 
$\alpha$-particle is then formed by fusing the two final deuterons. 

An intuitive motivation for this model can be obtained by looking at the
possible configurations of final particles. If the deuteron and 
$\alpha$-particle binding energies are neglected, the local 
$N\!N\rightarrow d\pi$ cm frames obviously coincide with the overall cm (CM) so 
that the two deuterons have the same energy, though not necessarily the same
directions. However, because of the strong $p$-wave dependence of the 
$N\!N\rightarrow d\pi$ reaction, the deuterons and pions are preferentially 
emitted at small angles to the beam direction. When deuterons are emitted 
parallel to each other, they are easily bound together into an 
$\alpha$-particle. Due to the kinematics, the associated produced pions then 
have small relative momenta and hence small invariant mass. This corresponds to 
the situation at the ABC peaks. In the second case of anti-parallel deuterons, 
the pions go back-to-back, their invariant mass is maximal and this is the 
reason for the central bump. However, it then needs a lot of Fermi momentum in 
the initial deuterons and/or final $\alpha$-particle in order to get the 
deuterons to stick together so that, despite the kinematical enhancement of this
configuration, it is suppressed compared to the parallel case. 

Both the ABC effect and the central bump are simultaneously explained in
our model by the strong $p$-wave dominance in the $N\!N\rightarrow d\pi$ 
reaction, with the former being helped by favorable kinematics.

\section{Formal description}
\label{sec:formal}

The considerations of the previous section are immediately transformed into the
Feynman diagram of Fig.~\ref{fig:ddapp}, where the momenta of the particles are
defined in the CM. In the following paragraphs the different parts of 
the corresponding matrix element are established using a description in terms 
of non-relativistic wave functions. The derivation of phase space formulae is
also given.

\subsection{Relativistic phase space}
\label{sec:phsprel}

The phase space calculations for the $dd\rightarrow\alpha\pi^{+}\pi^{-}$
reaction are performed relativistically starting from the general
expression~\cite{BD}

\begin{equation}
    {\rm d}\sigma = \frac{1}{|{\bf v}_{{\rm in}}|} \frac{1}{2E_{1}}
    \frac{1}{2E_{2}} |\mathcal{M}|^{2} 
    \frac{{\rm d}^{3}{k_{1}}{\rm d}^{3}{k_{2}}{\rm d}^{3}{k_{\alpha}}}
    {(2\pi)^{9}2\omega_{1}2\omega_{2}2\omega_{\alpha}}
    (2\pi)^{4}\,\delta^{4}(p_{1}+p_{2}-k_{1}-k_{2}-k_{\alpha})\:,
\label{eq;psdef}
\end{equation}
where $E_{i},{\bf p}_{i}$ and $\omega_{j},{\bf k}_{j}$
are the energies and momenta of the initial and final particles, respectively. 
The $\alpha$-particle is detected in the laboratory frame and thus 
$|{\bf v}_{{\rm in}}|=p/E$ and ${\rm d}^{3}{k_{\alpha}}/\omega_{\alpha}=
[k^{2}_{\alpha}\,{\rm d}\Omega_{\alpha}{\rm d} 
k_{\alpha}/\omega_{\alpha}]_{{\rm lab}}$, 
so that
\begin{equation}
\label{3.2}
    \left( \frac{{\rm d}^{2}\sigma}
    {{\rm d}\Omega_{\alpha}{\rm d} k_{\alpha}}\right)_{{\rm lab}}
 = 
    \frac{1}{32(2\pi)^{5}m_{d}\,p} 
    \frac{k_{\alpha{\rm lab}}^{2}}{\omega_{\alpha{\rm lab}}}
    \int \frac{{\rm d}^{3}k_{1}}{\omega_{1}} \frac{{\rm d}^{3}k_{2}}
    {\omega_{2}}|\mathcal{M}|^{2}
    \,\delta^{4}(p_{1}+p_{2}-k_{1}-k_{2}-k_{\alpha}).
\end{equation}

The remaining integrand in eq.~(\ref{3.2}) is relativistically invariant and the
resultant integral is most easily evaluated in the $\pi\pi$ rest frame, where 
the pions go back to back. Denoting with an asterisk quantities evaluated in
this frame, a straightforward calculation gives
\begin{equation}
    \mathcal{I} =  \int \frac{{\rm d}^{3}{k^{\ast}_{1}}}{\omega^{\ast}_{1}}
    \frac{{\rm d}^{3}{k^{\ast}_{2}}}{\omega^{\ast}_{2}} 
    |\mathcal{M}^{\ast}|^{2}\, \delta^{3}({\bf k}^{\ast}_{1}+
    {\bf k}^{\ast}_{2})
    \,\delta(M_X-\omega^{\ast}_{1}-\omega^{\ast}_{2}) 
    = \frac{k^{\ast}}{M_{X}}\int d\Omega^{\ast} 
    |\mathcal{M}^{\ast}|^{2}.
\label{eq:angint}
\end{equation}
The missing mass $M_{X}$ in the $dd\rightarrow\alpha{X}$ reaction is, in our
model, the effective mass of the pion-pion system $m_{\pi\pi}$. In terms of this
and ${\bf k}^{\ast}=\frac{1}{2}({\bf k}^{\ast}_{1}-{\bf k}^{\ast}_{2})$,
the differential cross section is
\begin{equation}
    \left( \frac{{\rm d}^{2}\sigma}
    {{\rm d}\Omega_{\alpha}{\rm d} k_{\alpha}}\right)_{{\rm lab}} =
    \frac{1}{32(2\pi)^{5}}\:
    \frac{k_{\alpha{\rm lab}}^{2}k^{\ast}}
        {m_{d}M_{X}p\,\omega_{\alpha{\rm lab}}}
    \int {\rm d}\Omega^{\ast}\: \frac{1}{9} 
    \sum_{{\rm spin,pol.}}|\mathcal{M}^{\ast}|^{2},
\label{eq:ph_sp_final}
\end{equation}

where the averaging over the initial spins and summing over the final is now 
included.

The description has so far concentrated on $\pi^{+}\pi^{-}$ production.
Simple isospin arguments suggest that $\pi^{0}\pi^{0}$ production should be
exactly half that of the charged pion but the narrowness of the ABC peak and the
significant pion mass differences make it necessary to evaluate the kinematics
separately in the two cases.

\subsubsection{Transformation to the $\pi\pi$ rest frame}
\label{sec:L_transf}

In order to implement the phase space considerations of the previous section, 
the matrix element has to be expressed in the $\pi\pi$ rest frame so that a
Lorentz transformation is needed between the CM and $\pi\pi$ systems.  Since in
general the $\alpha$-particle emerges at an arbitrary angle $\theta_{{\rm CM}}$
with respect to the beam direction, this involves a rotation. The righthanded CM
coordinate system is oriented with its $z$-axis in the beam direction 
(${\bf \hat{p}}$), the $x$-axis in the plane spanned by 
${\bf \hat{p}}$ and the $\alpha$-particle momentum (${\bf k}_{\alpha}$), 
and the $y$-axis parallel to  ${\bf \hat{p}\times\hat{k}}_{\alpha}$. 
A new coordinate system is introduced, by rotating about the $y$-axis, such that
the new $z$-axis coincides with the $\alpha$-particle direction. A Lorentz boost
is then applied in the $\alpha$-particle direction so that, in the new system, 
the pions emerge back to back. Explicitly this gives 
\begin{eqnarray}
    \frac{1}{2}(\omega_{1}-\omega_{2}) & = & 
    -\frac{k_{\alpha}k^{\ast}_{z'}}{m_{\pi\pi}} \\
    \frac{1}{2}(k_{1}^{x}-k_{2}^{x}) & = &
    k^{\ast}_{x'}\cos\theta_{{\rm CM}}+
    \gamma_{\pi}k^{\ast}_{z'}\sin\theta_{{\rm CM}} \\
    \frac{1}{2}(k_{1}^{y}-k_{2}^{y}) & = & k^{\ast}_{y'} \\
    \frac{1}{2}(k_{1}^{z}-k_{2}^{z}) & = &
    -k^{\ast}_{x'}\sin\theta_{{\rm CM}}+
    \gamma_{\pi}k^{\ast}_{z'}\cos\theta_{{\rm CM}},
\end{eqnarray}
where $xyz$ refer to the CM, and $x'y'z'$ to the $\pi\pi$ systems and
$\gamma_{\pi}=(\omega_{1}+\omega_{2})/m_{\pi\pi}$ is the 
Lorentz boost factor between the two.
The transverse and longitudinal components of the pion momentum difference
become

\begin{eqnarray}
    \half\,|{\bf k}_{1}^{b}\!-\!{\bf k}_{2}^{b}| 
    & = & \left[ (k^{\ast}_{x'}\cos\theta_{{\rm CM}}+
    \gamma_{\pi}k^{\ast}_{z'}\sin\theta_{{\rm CM}})^2+(k^{\ast}_{y'})^{2} 
    \right]^{\frac{1}{2}} \\
    \half\,(k_{1}^{z}-k_{2}^{z}) & = &
    -k^{\ast}_{x'}\sin\theta_{{\rm CM}},
    +\gamma_{\pi}k^{\ast}_{z'}\cos\theta_{{\rm CM}}
\end{eqnarray}
where $b$ indicates directions perpendicular to the beam.

\subsection{Vertex parametrizations}
\label{sec:vertpar}

\subsubsection{Split-up vertices}
Since large Fermi momenta are not required in our model, we shall only retain
the dominant $S$-state parts of the nuclear wave functions.
In a non-relativistic pole model the $d$:$pn$ vertices may be parametrized as:
\begin{equation}
    M_{d_{i}} =
    \left(\frac{-1}{\sqrt{2}}\bbox{\sigma}\cdot\bbox{\epsilon}_{i}\right)
    (2\pi)^{{3}/{2}}\frac{\sqrt{2m_{d}}}{m}
    ({\bf q}_{i}^{2}+\alpha_{d}^{2})
    \varphi({\bf q}'_{i})
\label{eq:dvert}
\end{equation}
where $\varphi({\bf q})$ is the deuteron $S$-state wave function in momentum 
space, ${\bf q}'_{i}\equiv ({\bf q}_{i}^{b},q_{i}^{z}/\gamma)$ the 
relative momentum of the nucleons, boosted in the $z$-direction to take the 
relativistic motion into account, and $\bbox{\epsilon}_{i}$ the deuteron 
polarization vector. The parameter $\alpha_{d}=\sqrt{mB_{d}}$, where
$m$ is the nucleon mass and ${B_d}$ the deuteron binding energy.

The $dd$:$\alpha$ vertex is parametrized in a similar manner by
\begin{equation}
    M_{\alpha} = 
    \left(\frac{-1}{\sqrt{3}}\bbox{\epsilon}\cdot\bbox{\epsilon}'\right)
    (2\pi)^{{3}/{2}} 2\sqrt{2m_{\alpha}}
    ({\bf q}_{\alpha}^{2}+\alpha^{2})
    \psi^{\dagger}({{\bf q}}_{\alpha}),
\end{equation}
where $\bbox{\epsilon}$ and $\bbox{\epsilon}'$ are the deuteron polarization 
vectors, $\psi({{\bf q}}_{\alpha})$ is the $\alpha$:$dd$ $S$-state wave 
function in momentum space, ${\bf q}_{\alpha}={\bf q}_{1}-{\bf q}_{2}-%
\frac{1}{2}({\bf k}_{1}-{\bf k}_{2})$ is the relative momentum of the 
deuterons, and $\alpha^2= m_{d}B_{\alpha}$, with $B_{\alpha}$ being the
$\alpha$-particle binding energy. Due to the large mass of the 
$\alpha$-particle and the relatively low pion momenta, there is no need to 
consider a Lorentz contraction factor in this case. 

\subsubsection{Pion production vertices}
The matrix element for the free $N\!N\rightarrow d\pi$ reaction is expressed
in terms of the partial wave amplitudes given in the appendix;
\begin{equation}
    M_{\pi_{i}} = K_{i}\mp\bbox{\sigma}\cdot\bbox{Q}_{i},
\end{equation}
where $K_{i}$ is the sum over all initial spin-singlet amplitudes and 
$\bbox{\sigma}\cdot{\bf Q}_{i}$ is the  sum over all triplet amplitudes. Note 
that, since we are now using $N\!N\rightarrow d\pi$ rather than 
$\pi d\rightarrow N\!N$, the deuteron polarization vectors should be 
charge-conjugated compared to those in the appendix. The $\mp$ signs for the 
triplet terms, referring to the upper and lower vertices of Fig.~\ref{fig:ddapp}
arise from the inversion of the direction of the incident nucleon momenta
in the two cases. The values of the amplitudes were calculated from the SAID 
database~\cite{SAID} using the normalization discussed in the appendix.  As seen
in Fig.~\ref{fig:ppdpi}, the nine amplitudes retained in the appendix reproduce 
the SAID predictions for the $\pi^{+}d\rightarrow{pp}$ differential cross 
section to within a few percent. For comparison the cross section 
calculated purely with the $^{1}\!D_{2}p$  amplitude, renormalized to fill the 
forward cross section, is also shown. This single amplitude, which was used in
our preliminary study~\cite{letter}, shows the manifest pion $p$-wave dominance
which is the main origin for the ABC enhancement in our model. 

In addition to considering far more input amplitudes, in the present analysis 
we retain also the explicit energy dependence prescribed by the database (see 
Sec.~\ref{sec:numerics}); in the earlier work we extracted the threshold $k$
factor and then assumed a constant reduced amplitude. The energy dependence 
of the $^{1}\!D_{2}p$ amplitude is hence different in the two analyses. 

\subsection{Matrix element}
\label{sec:matrix}

The matrix element deduced from the Feynman diagram of Fig.~\ref{fig:ddapp}
involves integration over two Fermi momenta
\begin{equation}
    \mathcal{M} = \int \frac{{\rm d}^{4}{q_{1}}}{(2\pi)^{4}}
        \frac{{\rm d}^{4}{q_{2}}}{(2\pi)^{4}} M_{{\rm tot}}\:,
\end{equation}
where $M_{{\rm tot}}$ is composed of the non-relativistic reduced vertex 
functions and the propagators stripped of their spin structure. In an order 
suggested by the diagram,

\begin{equation}
    M_{{\rm tot}} = 
    \sum_{{\rm int.\ spins}}
    M_{\alpha}P_{d}P_{d'} {\rm Tr}\left(
    M_{d_{2}}P_{2}^{c}M_{\pi_{1}}P_{4}
    M_{d_{1}}P_{3}^{c}M_{\pi_{2}}P_{1}\right),
\end{equation} 
where $P_{i}$ are the propagators, $M_{x}$ the matrix elements representing
the vertices and the sum is over the spins and polarizations of the internal
particles. Moreover, $c$ denotes charge conjugation.

\subsubsection{Approximation of propagators}
The denominators of the nucleon propagators are factorized to make the pole
structure explicit:
\begin{eqnarray}
    P_{1,4} & = & \frac{i\,2m}{(q_{20,10}+p_{0}-i\epsilon)
    [q_{20,10}-(\alpha_{d}^{2}
    +{\bf q}_{2,1}^{2}\pm{\bf p}\cdot{\bf q}_{2,1})/p_{0}+i\epsilon]} \\
    P^{c}_{2,3} & = & \frac{-i\,2m}{(q_{20,10}-p_{0}+i\epsilon)
    [q_{20,10}+(\alpha_{d}^{2}
    +{\bf q}_{2,1}^{2}\mp{\bf p}\cdot{\bf q}_{2,1})/p_{0}-i\epsilon]}.
\end{eqnarray}

Since the Fermi momenta (${\bf q}_{i}$) are generally quite small compared to
the incident deuteron momenta, only the poles closest to the origin at
$q_{10,20}=\frac{1}{p_{0}}(\alpha_{d}^{2}+
{\bf q}_{1,2}^{2}\mp{\bf p}\cdot{\bf q}_{1,2})$ yield significant
contributions. The integrations over $q_{20}$ and $q_{10}$ are then done by
contour integration in the lower half plane, yielding
\begin{equation}
    \int \frac{{\rm d} q_{20}\,{\rm d} q_{10}}
        {(2\pi)^{2}}\,P_{2}^{c}P_{4}P_{3}^{c}P_{1}
    \rightarrow \frac{-(2m)^{4}}{(2p_{0})^{2}
    ({\bf q}_{1}^{2}+\alpha_{d}^{2})
    ({\bf q}_{2}^{2}+\alpha_{d}^{2})} \sim
    \frac{-m^{2}}{\gamma^{2}({\bf q}_{1}^{2}+\alpha_{d}^{2})
    ({\bf q}_{2}^{2}+\alpha_{d}^{2})}\:,
\end{equation}
where the last step assumes that $p_{0}=E_{d}\sim 2\gamma m$. 

The deuteron propagators are
\begin{equation}
    P_{d,d'} = \frac{i}{Q_{d,d'}^{2}-m_{d}^{2}+i\epsilon},
\end{equation}
where $Q_{d,d'}=[p_{0}\pm(q_{10}-q_{20})
-\omega_{1,2},\pm({\bf q}_{1}-{\bf q}_{2})-{\bf k}_{1,2}]$.
Since ${\bf q}_{\alpha}^{2}+\alpha^{2}\sim -\frac{1}{2}
(Q_{d}^{2}-m_{d}^{2}+Q_{d'}^{2}-m_{d}^{2})$,
the product of the deuteron propagators and the $dd$:$\alpha$ vertex reduces to

\begin{equation}
    M_{\alpha}P_{d}P_{d'} = 
    \left(\frac{-1}{\sqrt{3}} \bbox{\epsilon}\cdot\bbox{\epsilon}'\right)
    (2\pi)^{\frac{3}{2}}
    \sqrt{2m_{\alpha}}
    \left( \frac{1}{Q_{d}^{2}-m_{d}^{2}+i\epsilon}+
        \frac{1}{Q_{d'}^{2}-m_{d}^{2}+i\epsilon}  \right)
    \psi_{\alpha}^{\dagger}\:.
\label{eq:alpha}
\end{equation}
The denominators in this expression can be expanded as
\begin{eqnarray}
    Q_{d,d'}^{2}-m_{d}^{2} & = & \left[ E_{d}-\omega_{1,
    2}\pm(q_{10}-q_{20}) \right]^{2}-
    \left[ \pm ({\bf q}_{1}-{\bf q}_{2})-{\bf k}_{1,2}
    -m_{d}^{2} \right]^{2} \nonumber \\
    && \sim \mp2m_{d} \left[ {\bf v}_{d}\cdot({\bf q}_{1}+{\bf q}_{2})+
    \frac{1}{2}(\omega_{1}-\omega_{2}) \right].
\label{eq:dpropden}
\end{eqnarray}
In the last step, the relation
\begin{equation}
    q_{10}-q_{20} = -\frac{1}{p_{0}} 
    [{\bf p}\cdot({\bf q}_{1}+{\bf q}_{2})
    -{\bf q}_{1}^{2}+{\bf q}_{2}^{2}] \sim 
    -{\bf v}_{d}\cdot({\bf q}_{1}+{\bf q}_{2})
\end{equation}
has been used, while quadratic terms were neglected.
Using these linearized forms, the principal value terms cancel in the sum of 
propagators in eq.~(\ref{eq:alpha}) to leave only a $\delta$-function term;
\begin{equation}
    M_{\alpha}P_{d}P_{d'} = 
    \left( \frac{-1}{\sqrt{3}} \bbox{\epsilon}\cdot\bbox{\epsilon}' \right)
    (2\pi)^{\frac{3}{2}}
    \sqrt{2m_{\alpha}}\psi_{\alpha}^{\dagger}
    \frac{-i\pi}{m_{d}v_{d}}
    \delta\!\left( q_{1}^{z}+q_{2}^{z}+
        \frac{\omega_{1}-\omega_{2}}{2v_{d}} \right)\:.
\end{equation}

Assuming that Fermi momentum effects may be neglected in the spin couplings, 
which is a good approximation in the ABC peak regions, the matrix element is 
written as the product
\begin{equation}
    \mathcal{M} \equiv -i\frac{m_{\alpha}}{v_{d}}\mathcal{W\,K},
\end{equation}
where the spin kernel $\mathcal{K}$ and dimensionless scalar form factor 
$\mathcal{W}$ are given, respectively, by
\begin{equation}
    \mathcal{K} \equiv  
    \sum_{{\rm int.\ spins}}
    \frac{-1}{\sqrt{3}}\bbox{\epsilon}\cdot\bbox{\epsilon}'
    {\rm Tr} \left[ \left( \frac{-\bbox{\sigma}\cdot\bbox{\epsilon}_{2}}
    {\sqrt{2}} \right)
    \left( K_{1}-\bbox{\sigma}\cdot{\bf Q}_{1} \right)
    \left( \frac{-\bbox{\sigma}\cdot\bbox{\epsilon}_{1}}{\sqrt{2}} \right)
    \left( K_{2}+\bbox{\sigma}\cdot{\bf Q}_{2} \right) \right]\:,
\label{eq:Kdef}
\end{equation}
and
\begin{eqnarray}
\nonumber
    \mathcal{W} & \equiv & \frac{1}{\sqrt{\pi m_{\alpha}}}
    \int \frac{{\rm d}^{3}{q_{1}}{\rm d}^{3}{q_{2}}}{\gamma^{2}}
    \varphi_{d}({\bf q}_{1}')
    \varphi_{d}({\bf q}_{2}')
    \psi_{\alpha}^{\dagger}({\bf q}_{\alpha})
    \,\delta\!\left( q_{1}^{z}+q_{2}^{z}+
        \frac{\omega_{1}-\omega_{2}}{2v_{d}} \right)\:\cdot\\
\label{eq:Wdef}
\end{eqnarray}

\subsection{Summing over spins and polarizations}
\label{sec:spinsum}
The spin kernel of eq.~(\ref{eq:Kdef}) can be simplified somewhat by expanding 
the product and taking the trace. This results in the spin-decoupled form
\begin{equation}
    \mathcal{K} = \frac{-1}{\sqrt{3}}
    \left[ \mathcal{A}(\bbox{\epsilon}_{1}\cdot\bbox{\epsilon}_{2})
    -i\overline{\mathcal{B}}\cdot(\bbox{\epsilon}_{1}\bbox{\times\epsilon}_{2})
    -\bbox{\epsilon}_{1}\cdot\overline{\overline{\mathcal{C}}}
    \cdot\bbox{\epsilon}_{2} \right],
\end{equation}
where the spin-0, -1, and -2 amplitudes are
\begin{eqnarray}
    \mathcal{A} & \equiv & \sum_{{\rm pol.}} 
    (\bbox{\epsilon}\cdot\bbox{\epsilon}')
    \left( K_{1}K_{2}+\fmn{1}{3}{\bf Q}_{1}\cdot{\bf Q}_{2} \right) \\
    \overline{\mathcal{B}} & \equiv & \sum_{{\rm pol.}} 
    (\bbox{\epsilon}\cdot\bbox{\epsilon}')
    \left( K_{1}{\bf Q}_{2}+K_{2}{\bf Q}_{1} \right) \\
    \overline{\overline{\mathcal{C}}} & \equiv & \sum_{{\rm pol.}} 
    (\bbox{\epsilon}\cdot\bbox{\epsilon}') \left(
    {\bf Q}_{1}{\bf Q}_{2}+{\bf Q}_{2}{\bf Q}_{1}-
    \fmn{2}{3}{\bf Q}_{1}\cdot{\bf Q}_{2} \right)
\end{eqnarray}
and the sums are over the polarizations of the internal deuterons.

The unpolarized matrix element squared is then
\begin{equation}
    \sum_{{\rm ext.\ pol.}}|\mathcal{K}|^{2} =
    \fmn{1}{3} \left( 3|\mathcal{A}|^{2}+2|\overline{\mathcal{B}}|^{2}
    +|\overline{\overline{\mathcal{C}}}|^{2}
    \right),
\label{eq:k2}
\end{equation}
where $|\overline{\overline{\mathcal{C}}}|^{2} \equiv
\sum_{i,j}\mathcal{C}_{ij}\mathcal{C}_{ij}^{\dagger}$ and the sum is over the
polarizations of the external deuterons. The deuteron vector and tensor 
analyzing powers $A_{y}$ and $A_{yy}$ are obtained from the sums
\begin{equation}
    \sum_{i,j,k}\mathcal{K}_{ij}P_{jk}\mathcal{K}_{ik}^{\dagger},
\label{eq:Kijdef}
\end{equation}
where the $P_{ij}$ are the spin projection operators in a Cartesian 
basis~\cite{Madison}. Since the dominant $^{1}\!D_{2}p$ amplitude by itself
gives only a contribution to the $\mathcal{A}$ amplitude, this would lead to
$A_{y}=A_{yy}=0$. The analyzing powers are therefore very sensitive probes of
the importance of non-dominant $N\!N\rightarrow d\pi$ input amplitudes.

To avoid the tedious algebra associated with nine partial wave amplitudes
occurring bilinearly in each {\it amplitude}, the above sums and contractions
were in practice carried out numerically.

\subsection{Form factor in configuration space}
\label{sec:formfact}

The form factor of eq.~(\ref{eq:Wdef}) is most easily evaluated by 
transforming to configuration space;

\begin{eqnarray}
    \mathcal{W} & = & \frac{1}{\sqrt{2m_{\alpha}}} \int
    {\rm d}^{2}b\,{\rm d} z_{1}{\rm d} z_{2}
    \,\Phi_{d}({{\bf b}},\gamma z_{1})
    \Phi_{d}({{\bf -b}},\gamma z_{2})
    \Psi_{\alpha}^{\dagger}\!({\bf x}_{\alpha})
\nonumber \\
    & \times & \exp\!\left[ -\frac{i}{2}
    ({\bf k}_{1}-{\bf k}_{2})\cdot{\bf x}_{\alpha} \right]\,
    \exp\!\left[ -\frac{i}{4v_{d}}(\omega_{1}-\omega_{2})(z_{1}+z_{2})
    \right]\:,
\end{eqnarray}
where ${\bf x}_{\alpha}=({\bf b},(z_{1}\!-\!z_{2})/2)$.

Since the wave functions have been taken to be spherically symmetric, the
angular integration over the $b$ variable can be performed explicitly.
Furthermore, only the even parts of the exponentials will contribute, so that 
\begin{eqnarray}
    \lefteqn{\mathcal{W} = \frac{2\pi}{\sqrt{2m_{\alpha}}} \int
    b\,{\rm d} b\,{\rm d} z_{1}{\rm d} z_{2}\,
    \Phi_{d}({{\bf b}},\gamma z_{1})\,
    \Phi_{d}({{\bf -b}},\gamma z_{2})\,
    \Psi_{\alpha}^{\dagger}\!\left({{\bf b}},\frac{z_{1}\!-\!z_{2}}{2}\right) }
    \nonumber \\[1ex]
    & \times &
    J_{0}\!\left(\half|{\bf k}_{1}^{b}\!-\!{\bf k}_{2}^{b}|b\right)
    \cos\!\left[ \fmn{1}{4}(k_{1}^{z}-k_{2}^{z})(z_{1}-z_{2}) \right]
    \cos\!\left[ \fmn{1}{4v_{d}}(\omega_{1}-\omega_{2})(z_{1}+z_{2})
    \right]. 
\label{eq:ffdef}
\end{eqnarray}

\subsection{Reduction of cross sections due to flux damping}
\label{sec:dist}
The fluxes of initial and final particles are reduced because of multiple
scattering. Any detailed microscopic estimate of such effects would rest upon 
assumptions that are difficult to justify quantitatively. For example, in any 
Glauber-like approximation, correlation effects would be very large for the 
light nuclei that we are considering. We therefore limit our ambition, 
attempting a crude estimate through an overall scale factor. This should at 
least give a hint of the magnitude and direction of the damping. Since our model
is based upon two $N\!N\rightarrow{d}\pi$ reactions, it is natural to estimate 
the reduction in flux through the factor
\begin{equation}
    \mathcal{D} =
    \frac{\sigma_{{\rm tot}}(dd)}
    {2[\sigma_{{\rm tot}}(pn)+\sigma_{{\rm tot}}(pp)]}\times
    \left(
    \frac{\sigma_{{\rm tot}}(\pi\alpha)}{2\sigma_{{\rm tot}}(\pi{d})}
    \right)^2,
\label{eq:dist}
\end{equation}
where the last factor is squared since there are two emerging pions.

The assumption behind this formula is that in the initial state the flux 
reduction due to multiple scattering is the same in all $dd$ reaction. A 
similar argument is valid in the final state, assuming the $\alpha$-particle to
be a $dd$ aggregate. The values of total cross sections needed were taken from
\cite{PDG}~($N\!N$), \cite{Kishida}~($dd$), \cite{Pedroni}~($\pi{d}$), and
\cite{pialpha}~($\pi\alpha$). Interpolation between different data sets was 
often needed and the values used are collected in Table~\ref{tab:dist}, together
with the calculated reduction $\mathcal{D}$. The latter is compared to the 
factor needed to scale our model calculations to fit the data, as shown in the 
figures. While our simplistic estimate goes in the right direction, it does not
account for all the deviations from the data. There is however considerable 
uncertainty in the pion distortion at the lower energies.

\section{Numerical calculation}
\label{sec:numerics}

The evaluation of cross sections and polarization observables was done in two 
steps. The form factor of eq.~(\ref{eq:ffdef}) was firstly obtained using 
standard Gauss-Legendre routines with 50 points in each of the three 
dimensions and tabulated in steps of the parameters $k_{\alpha{\rm lab}}$, 
$\varphi^{\ast}$, and $\cos\theta^{\ast}$. In the second step the spin sums and
contractions were done at each of these points, followed by the integration
over phase space, including the tabulated form factor.

\subsection{Wave functions}
The $S$-state Paris~\cite{Paris} wave function is used to describe the 
deuterons, whereas the $\alpha$-particle is represented by the $dd$:$\alpha$ 
cluster function of Forest et al.~\cite{dumbbell}. It is shown in the latter 
work that the nucleon-nucleon distributions in the $\alpha$-particle and the 
deuteron are very similar for short distances ($r<2~$fm), with a constant scale
factor of 4.7 between them. Since pion absorption on deuterons occurs mainly 
when the nucleons are close together, we need the number $N_{\alpha}$ of pairs
of such ``small'' deuterons in the $\alpha$-particle. Assuming the
neutron-proton distribution inside the two deuterons of the $\alpha$-particle to
be independent of each other, we therefore normalize the wave function to
$N_{\alpha}=2.3$.

\begin{figure}[p]
\centering    \includegraphics{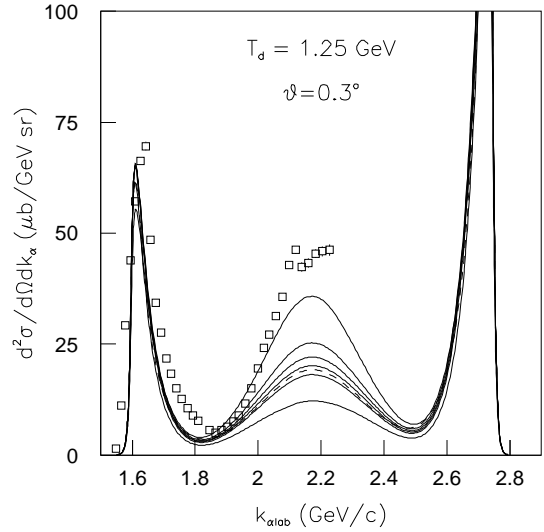}
\caption{Comparison of cross sections calculated with different
$\alpha$-particle wave functions. Solid lines correspond to the Gaussian wave
function with the parameter $\beta$ (defined in the text) having the values 
(from bottom to top): 0.55, 0.64, 0.67, 0.70, 0.75, and 1.0 ${\rm fm}^{-1}$. 
For comparison the calculation with the Forest~et~al.~\protect\cite{dumbbell} 
wave function (dashed line), is repeated. The data are from 
Ref.~\protect\cite{Ban76}.}
\label{fig:wfgauss}
\end{figure}

Since the shape of the $\alpha$:$dd$ function is similar to a Gaussian,
$\exp(-\beta^2r^2)$, it is easy to test our dependence on it by varying the 
parameter $\beta$. The result of this test (using the constant amplitudes of 
next section) at $T_{d} = 1250$~MeV and $\theta_{\alpha}=0.3^{\circ}$ is shown 
in Fig.~\ref{fig:wfgauss}. While the ABC peaks remain largely unaffected, there
is large sensitivity in the central region. Calculations at $11.0^{\circ}$ 
and also at 787~MeV show less variation. Hence this uncertainty is associated 
with high missing masses.

\subsection{Partial wave amplitudes}
The values of the partial wave amplitudes can be extracted from the SAID
database~\cite{SAID} in several ways. The first and simplest is to use constant
amplitudes evaluated at the kinematics corresponding to the edges of the 
spectra, i.e.\, close to the ABC peaks, where the two pions have 
identical momenta. The single amplitude used in our previous work
was extracted in this way~\cite{letter}.

Energy dependent input --- where the amplitudes are calculated separately for
each point of integration --- is an attempt to take the Fermi motion in the 
$\alpha$-particle into account. There is no unambiguous way to do this and so 
we have adopted two slightly different methods.  One based on the previous 
technique, but calculated at each point separately,  will be called the E1 
prescription. The second approach (E2) uses the same pion  momenta but now 
assumes the deuterons to have half the $\alpha$-particle momentum.  The two 
clearly coincide at the edges of the missing mass spectra, but have different 
characteristics in the central region. Their relative merits will be discussed 
in the next section.

\section{Results}
\label{sec:results}

\begin{figure}[p]
\mbox{}\hspace*{-15mm}\includegraphics{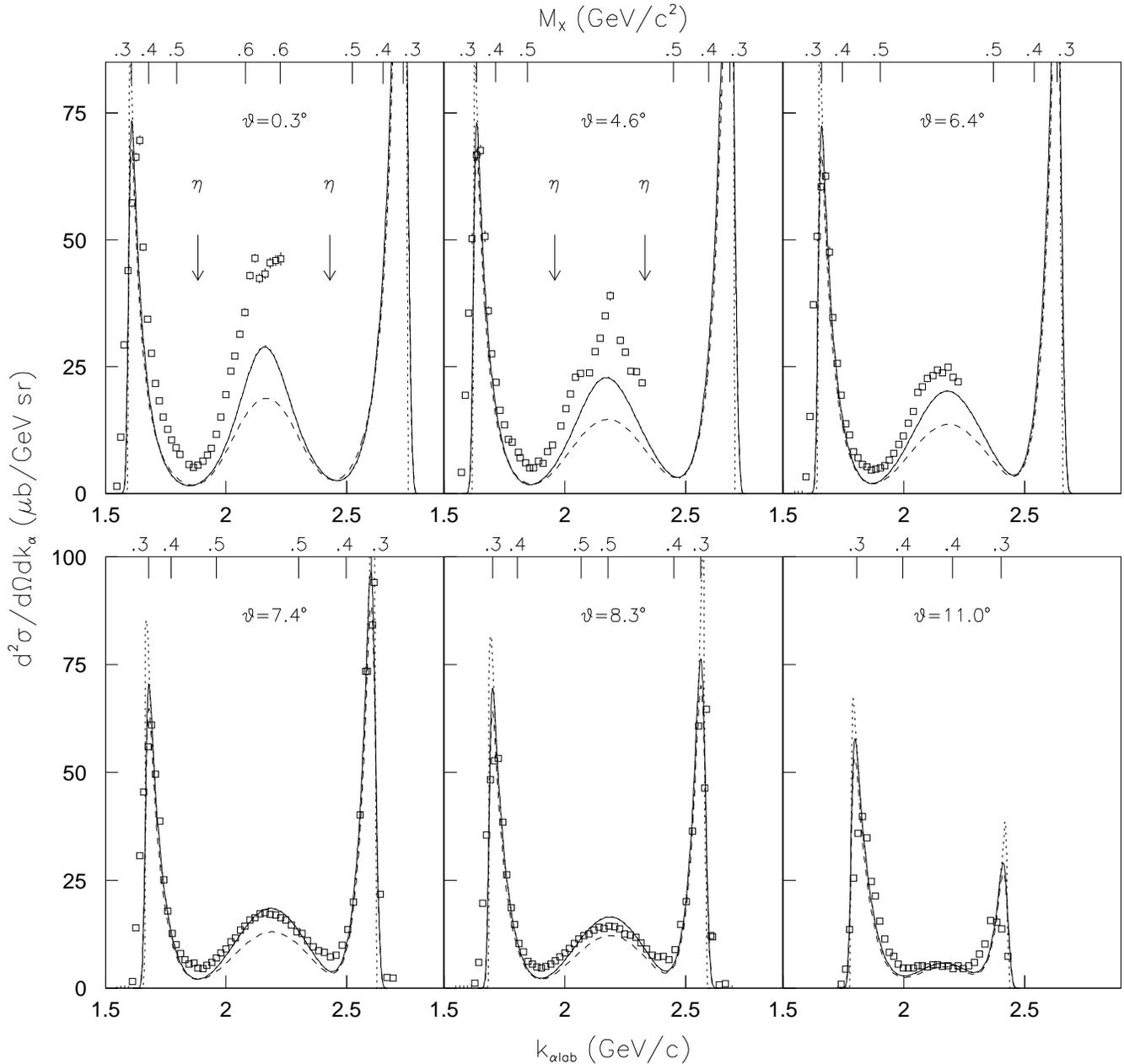}
\caption{Angular distribution $dd\rightarrow\alpha{X}$ at $T_{d}=$~1.25~GeV. 
The smeared E1 (dashed line) and E2 (solid line) calculations are compared to 
the data of Ref.~\protect\cite{Ban76}. For E2 the raw calculations (dotted line)
is also given. Our predictions are fitted to the forward (low $k_{\alpha}$) ABC
peak with a scale factor 0.71, cf. Sec.~\ref{sec:dist}.}
\label{fig:Ban_1250}
\end{figure}

\begin{figure}[p]
\mbox{}\hspace*{-15mm}\includegraphics{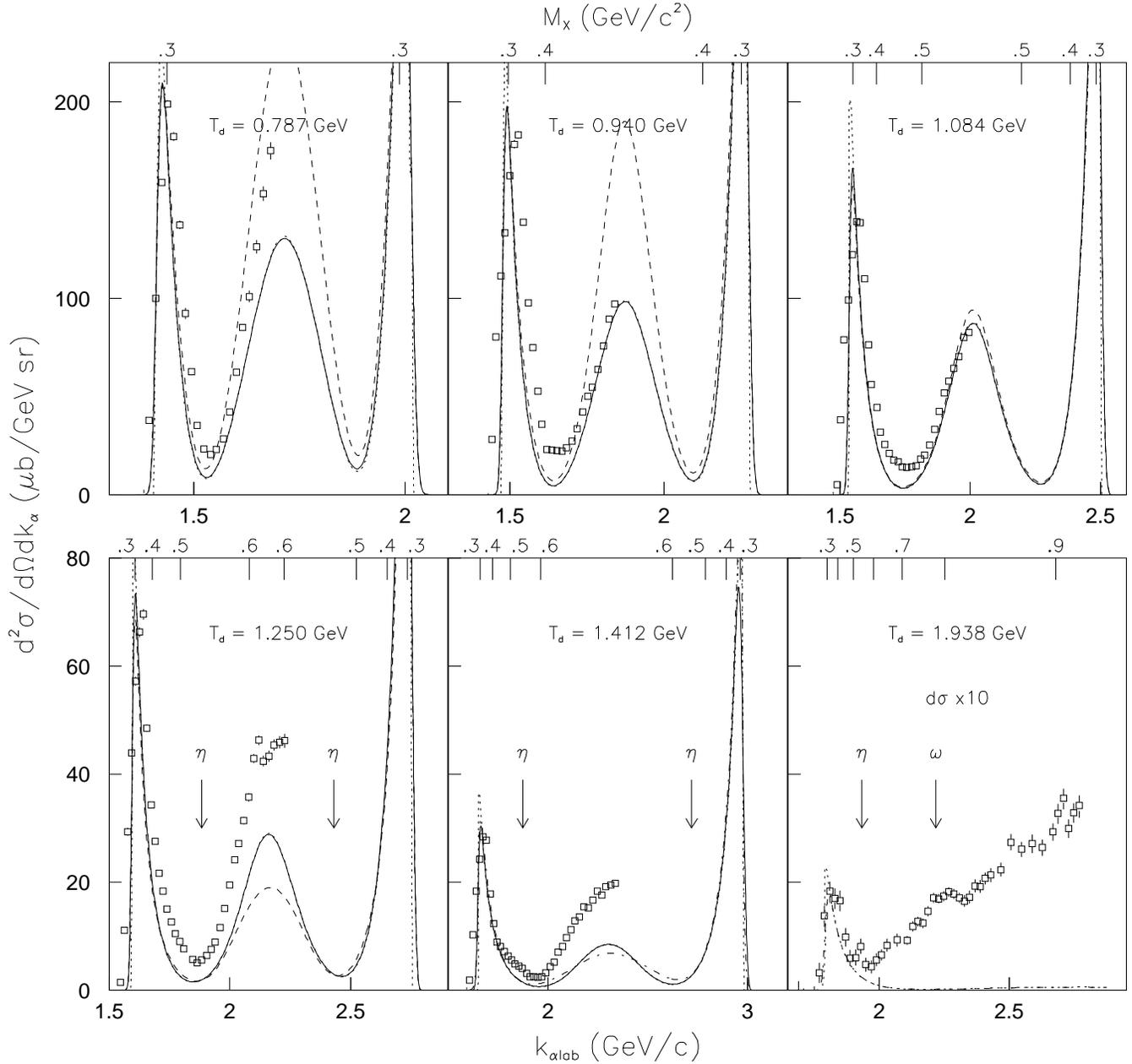}
\caption{Predictions of the energy dependence of $dd\rightarrow\alpha{X}$ at
$\theta_{\alpha}=0.3^{\circ}$ compared to data from Ref.~\protect\cite{Ban76}. 
The different curves are defined as in Fig.~\ref{fig:Ban_1250}. Since it is 
impossible to use E1 at 1.4~GeV, and neither E1 nor E2 at 1.9~GeV, the 
calculations with constant amplitudes (dot-dashed line) are given at these 
energies. The calculations are reduced by the scale factors given in 
Sec.~\ref{sec:dist}.}
\label{fig:Ban_all_en}
\end{figure}

The predictions of our model are compared with the Saclay unpolarized cross 
section data of Ref.~\cite{Ban76} in Figs~\ref{fig:Ban_1250} and
\ref{fig:Ban_all_en}. Shown are the raw results from the E2 prescription and
also those obtained by smearing E1 and E2 over a Gaussian experimental
resolution in the $\alpha$-particle momentum with a relative standard deviation
$\sigma(k_{\alpha})/k_{\alpha}=0.5\%$~\cite{Ban76}. The calculations are scaled
to fit the data with factors listed in Table~\ref{tab:dist}. This scaling is
independent of the choice of amplitudes. The agreement with data is remarkably 
good, both in angular distribution and energy dependence. The differences found
between the scale factors and the flux damping estimates could be attributed 
mainly to the crudeness of the estimate and the uncertainties in its input data.

The slight discrepancy at the central bump for small angles could be due to the 
production of three pions or to our simplified treatment of the Fermi momenta. A
broad $\sigma$-meson~\cite{PDG} would also enhance the cross section here. An 
estimate of the possible contribution from production of $\eta$-mesons could be
obtained from a measurement at 1.95~GeV~\cite{Ban85}. The cross section is 
however so low ($\sim 1~\mu{\rm b/(sr GeV/c)}$) that it should be invisible in
all the plots, except possibly for 1.94~GeV where one might see a hint of a 
peak. As illustrated in Fig.~\ref{fig:wfgauss}, the central region is also 
sensitive to the precise form assumed for the $\alpha$-particle wave function. 
Inclusion of $D$-states in the wave functions would give effects for large 
missing mass as well since it corresponds to large Fermi momenta.

Apart from a small change in the ABC peaks, the main difference between E1 and 
E2 is found in the central region. The reason for this is the different 
kinematics used in the two cases --- E1 tends to overstate the $\pi$d energy in 
the central bump and thus attains the maximal $N\!N\rightarrow d\pi$ cross 
section already at 940~MeV, while E2 has its maximum at 1250~MeV. This accounts 
for the E1 overestimating (compared to E2) at low missing masses and 
underestimating at high maximal missing mass. At 1250~MeV and $11.0^{\circ}$ the
missing mass is not high enough to be sensitive to the difference between the 
two assumptions. Unfortunately the range of data in the SAID 
database~\cite{SAID} does not extend to the energies needed for E1 at 1412~MeV 
and for E2 at 1938~MeV. The close agreement between E2 and the calculation with 
constant amplitudes at 1412~MeV justifies the use of constant amplitudes at the
higher energies in Fig.~\ref{fig:Ban_all_en}.

The underestimate at 787~MeV is probably an indication of a more severe problem
within our double-$\Delta$ model. There are two measurements at even lower
energies, one close to threshold at $T_{d}=$~570~MeV~\cite{Barg} and another at
650~MeV~\cite{Chap}. In both cases the data are structureless, and consistent 
with pure phase space, with total cross sections of $\sigma(570)\sim 43$~nb and
$\sigma(650)\sim 600$~nb, respectively. In our model a pronounced ABC-like 
structure should be clearly visible at both energies. However, our predicted 
total cross sections are a factor of about 20 too low and it therefore seems 
that there is some other mechanism which is mainly responsible for the smooth 
behavior near threshold. This extra contribution might account already for our 
underestimate at 787~MeV.

\begin{figure}[p]
\mbox{}\hspace*{25mm}\includegraphics{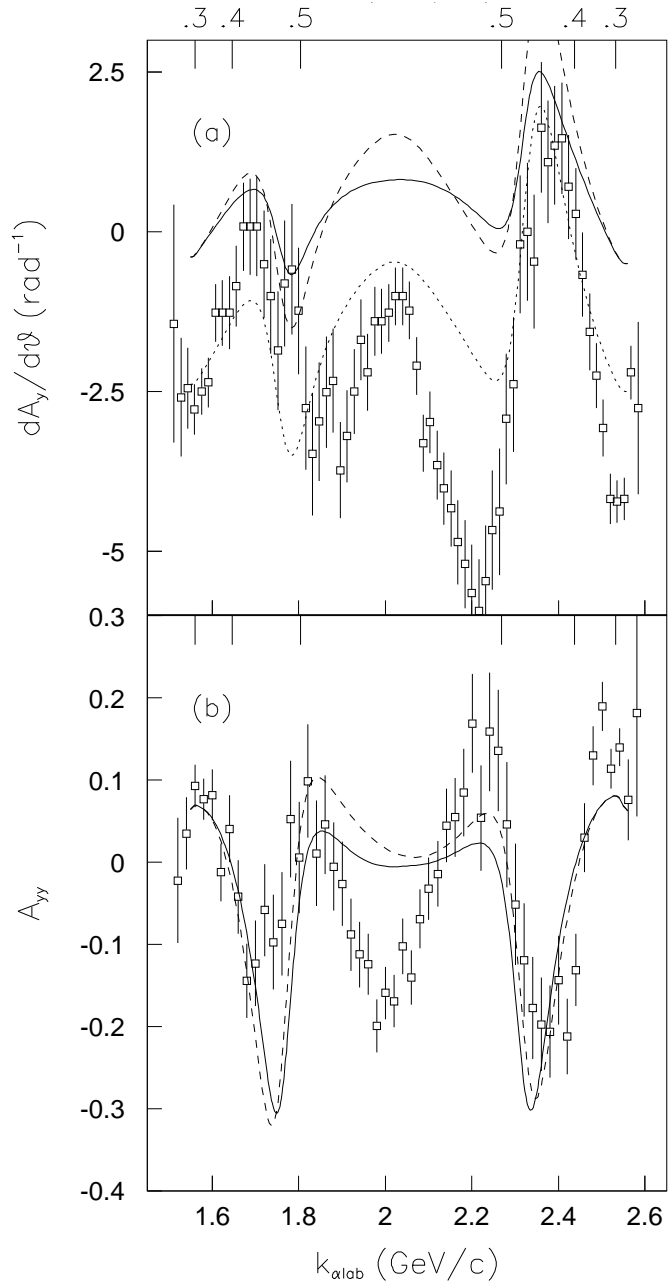}
\caption{Predictions of forward analyzing powers compared to the SPESIII
measurements~\protect\cite{SPESIII}. Both the E1 (dashed line) and E2 
(solid line) calculations are given. The experimental acceptance is
incorporated in the theoretical results.
{\bf (a)} Average slope of vector analyzing power $A_{y}'$. The similarity in
shape is more evident after shifting the E1 prediction by 
$-2\ {\rm rad}^{-1}$ (dotted line). Data are obtained for a range of
energies 1.116 $<T_{d}<$ 1.127~GeV, while calculations are done at
1.122~GeV. {\bf (b)} Tensor analyzing power $A_{yy}$ at $T_{d}=1.116$~GeV.}
\label{fig:ayyay_art}
\end{figure}

The deuteron analyzing powers measured by the SPESIII spectrometer at 
Saturne~\cite{SPESIII} are seen in Fig.~\ref{fig:ayyay_art} together with our 
model predictions integrated over the spectrometer acceptance. The slope of 
vector analyzing power $A_{y}'$ in the forward direction is calculated according
to the experimental average procedure described in Ref.~\cite{SPESIII}. The 
differences between the E1 and E2 calculations are clearly seen for $A_{y}'$, 
while they are rather marginal for $A_{yy}$. This greater sensitivity of 
$A_{y}'$ could be due to its dependence upon the relative phase between singlet
and triplet amplitudes and also the importance of the strongly energy-dependent
and dominant $^{1}\!D_{2}p$ amplitude. The E1 predictions are shifted by an 
arbitrary $-2\ {\rm rad}^{-1}$ in the figure to show the close resemblance to 
the data structures. Any small extra terms in the amplitude could cause such a 
shift. For both analyzing powers, there are again discrepancies in the central 
region, presumably for the same reasons as for the unpolarized cross sections.

All calculations were done using the C500 solution of~\cite{SAID} but, in order
to check the stability of the result, they were repeated for the SP96 solution
as well. The two sets of results are very close and the small differences found
are completely overshadowed by the larger uncertainties inherent in the 
procedures for extracting amplitudes.

\section{Conclusions}
\label{sec:concl}

For the first time a quantitative model has been proposed for the ABC effect in
the $dd\rightarrow\alpha X$ reaction. Despite its simplicity, and without
benefitting from any free parameters, it is able to reproduce all the main 
features of the $\alpha$-particle momentum spectrum observed throughout the 
double-$\Delta$ region. It is hence clear that, at least for this reaction, the
ABC effect is indeed a kinematical enhancement in the independent production of
two $p$-wave pions when these emerge with parallel momenta. 

The analyzing power predictions are equally impressive and reproduce quite
nicely the frequency and strength of the oscillations for both $A_{yy}$ and
$A_{y}'$. Since these quantities are sensitive to the non-dominant input 
amplitudes, and in particular their relative phases, this more thorough 
examination of the details of our model adds further strength to the conclusion 
that it supplies the correct dynamics in the double-$\Delta$ region.

The low energy data present, however, a more delicate situation to interpret.
We do not think that it is possible to force the present model to reproduce
pure phase space at low energies. Since the $p$-wave dominance of 
$N\!N\rightarrow d\pi$ extends almost down to threshold, the $s$-wave pions 
needed to furnish isotropy could not be produced via that subprocess. The 
analogous contribution of isovector nucleon-nucleon pairs, in for example 
$pp\rightarrow pp\pi^0$, is known to be much smaller than isoscalar pairs both 
near threshold and throughout the $\Delta$-peak~\cite{Hahn}.  There must 
therefore be another production mechanism present, which is dominant at low 
energies but which assumes lesser importance at higher energies where the ABC is
seen clearly. In order to improve our understanding of the dynamics in this
kinematical region, we suggest that further experiments should be done at 
deuteron beam energies $T_{d}<$~0.8~GeV. In addition to inclusive polarized and 
unpolarized cross sections, exclusive reactions are also needed so that angular 
distributions of pairs of the final particles can be constructed. Such data
already exist for $pd\rightarrow\,^3$He$\,\pi^{+}\pi^{-}$~\cite{MOMO}.
Since all such quantities can be calculated within our model they provide extra
confirmations and tests of it.

Given its quantitative success it is natural to ask if our model could be 
extended to include other reactions as well. A simple extension would be to 
calculate the double photon production $dd\rightarrow\alpha\gamma\gamma$ via two
$np\rightarrow{d}\gamma$ reactions. This could be an important background in the
measurement of the charge-symmetry-breaking $dd\rightarrow\alpha\pi^{0}$ 
reaction, observed at Saturne in the double-$\Delta$ region~\cite{Goldzahl}. We
are currently investigating this possibility.

\section*{Acknowledgement}
We would like to thank Pia Th\"orngren Engblom of the Stockholm Nuclear Physics
Group whose experiment at the Celsius ring~\cite{Barg} was the starting point
for this study. This work has been made possible by the continued financial 
support of the Swedish Royal Academy and the Swedish Research Council, and one 
of the authors (CW) would like to thank them and The Svedberg Laboratory for 
their generous hospitality.

\appendix
\newpage

\section{Partial wave amplitudes and Clebsch-Gordan coefficients expressed by
 spin amplitudes}
\label{app:pwa}
\setcounter{equation}{0}

We find it simpler to estimate the double-$\Delta$ contribution to the two-pion
production using a spin-amplitude description of the $\pi^{+}d\rightarrow pp$
reaction rather than employing partial waves. We therefore give here expressions
for the lower partial waves in terms of the Pauli spin matrices 
($\bbox{\sigma}$), the deuteron polarization vectors ($\bbox{\epsilon}$), and 
the proton and pion momenta (${\bf p}$ and ${\bf k}$). The partial waves are 
denoted by $^{2S_{pp}+1}(L_{pp})_{J}l_{\pi}$, where $S_{pp}$ and $L_{pp}$ are 
the spin and orbital angular momentum of the proton-proton system in the initial
state, $J$ is the total angular momentum, and $l_{\pi}$ is the pion angular 
momentum in the final state. The operators $\mathcal{O}_i$ corresponding to 
particular partial wave transitions are given below in the normalization $\int
{\rm d}\Omega\sum_{\bbox{\epsilon}}{\rm Tr}|\mathcal{O}_{i}|^2=
2\cdot4\pi (2J_i+1)$.

\subsection{Singlet amplitudes}
\begin{eqnarray}
\mathcal{O}\left(^{1}\!S_{0}p\right) & = & 
        -{\bf \hat{k}}\cdot\bbox{\epsilon} \\
\mathcal{O}\left(^{1}\!D_{2}p\right) & = & \sqrt{\frac{5}{2}} \left\{
    3({\bf \hat{p}}\cdot{\bf \hat{k}})
({\bf \hat{p}}\cdot\bbox{\epsilon})-{\bf \hat{k}}\cdot\bbox{\epsilon} 
    \right\} \\
\mathcal{O}\left(^{1}\!D_{2}f\right) & = & \frac{\sqrt{15}}{2} \left\{
    2({\bf \hat{p}}\cdot{\bf \hat{k}})
    ({\bf \hat{p}}\cdot\bbox{\epsilon})-
    \left[ 5({\bf \hat{p}}\cdot{\bf \hat{k}})^{2}-1\right] 
    {\bf \hat{k}}\cdot\bbox{\epsilon}
    \right\} \\
\mathcal{O}\left(^{1}\!G_{4}f\right) & = & \frac{3}{4}\left\{ 
    35({\bf \hat{p}}\cdot{\bf \hat{k}})^{3}
    ({\bf \hat{p}}\cdot\bbox{\epsilon})
    +3({\bf \hat{k}}\cdot\bbox{\epsilon}) \right.
    \nonumber \\
    & - & 15 \left. \left[ 
    ({\bf \hat{p}}\cdot{\bf \hat{k}})^{2}
    ({\bf \hat{k}}\cdot\bbox{\epsilon})+
    ({\bf \hat{p}}\cdot{\bf \hat{k}})
    ({\bf \hat{p}}\cdot\bbox{\epsilon}) \right]
    \right\}
\end{eqnarray}

\subsection{Triplet amplitudes}
\begin{eqnarray}
\mathcal{O}\left(^{3}\!P_{1}s\right) & = & \sqrt{\frac{3}{2}}i
    \bbox{\epsilon}\cdot({\bf \hat{p}}\bbox{\times\sigma})
 \\
\mathcal{O}\left(^{3}\!P_{1}d\right) & = & -\frac{\sqrt{3}}{2}i \left\{
    3({\bf \hat{k}}\cdot\bbox{\epsilon})
    {\bf \hat{k}}\cdot({\bf \hat{p}}\bbox{\times\sigma})
    -\bbox{\epsilon}\cdot({\bf \hat{p}}\bbox{\times\sigma}) \right\}
 \\
\mathcal{O}\left(^{3}\!P_{2}d\right) & = & -\frac{\sqrt{15}}{2}i \left\{
    ({\bf \hat{p}}\cdot{\bf \hat{k}})
    \bbox{\sigma}\cdot({\bf \hat{k}}\bbox{\times\epsilon})
    -(\bbox{\sigma}\cdot{\bf \hat{k}})
    {\bf \hat{k}}\cdot({\bf \hat{p}}\bbox{\times\epsilon}) \right\}
 \\
\mathcal{O}\left(^{3}\!F_{2}d\right) & = & \sqrt{\frac{5}{2}}i \left\{
    5({\bf \hat{p}}\cdot{\bf \hat{k}})
    ({\bf \hat{p}}\cdot\bbox{\sigma})
    {\bf \hat{p}}\cdot({\bf \hat{k}}\bbox{\times\epsilon})
    \right. \nonumber \\
    & - & \left.
    \left[({\bf \hat{p}}\cdot{\bf \hat{k}})
    \bbox{\sigma}\cdot({\bf \hat{k}}\bbox{\times\epsilon})
    +(\bbox{\sigma}\cdot{\bf \hat{k}})
    {\bf \hat{p}}\cdot({\bf\hat{k}}\bbox{\times\epsilon}) 
    \right] \right\}
 \\
\mathcal{O}\left(^{3}\!F_{3}d\right) & = & \frac{\sqrt{7}}{4}i \left\{
    \left[ 5({\bf \hat{p}}\cdot{\bf \hat{k}})^{2}-1 \right]
    \bbox{\epsilon}\cdot({\bf \hat{p}}\bbox{\times\sigma}) \right. 
    \nonumber \\
    & + & \left. 10({\bf \hat{p}}\cdot{\bf \hat{k}})
    ({\bf \hat{p}}\cdot\bbox{\epsilon})
    {\bf \hat{k}}\cdot({\bf \hat{p}}\bbox{\times\sigma})
    -2({\bf \hat{k}}\cdot\bbox{\epsilon})
    {\bf \hat{k}}\cdot({\bf \hat{p}}\bbox{\times\sigma}) \right\}
\end{eqnarray}

\subsection{Clebsch-Gordan coefficients}

In the $\bbox{\sigma}$-$\bbox{\epsilon}$ expressions for Clebsch-Gordan 
coefficients, the non-relativistic two-component spinor $\eta$ and its 
charge conjugate $\eta_{c}\equiv-i\sigma^{2}\eta^{\ast}$ are introduced.

\begin{eqnarray}
    \left\langle 1\lambda \left| 
    \frac{1}{2}m_{1} \right. \frac{1}{2}m_{2} \right\rangle & \leftrightarrow & 
    \eta_{c1}^{\dagger}
    \left( \frac{-1}{\sqrt{2}}\,
    \bbox{\sigma}\cdot\bbox{\epsilon}^{(\lambda)\dagger} \right)
    \eta_{2} \\
    \left\langle \frac{1}{2}m_{1} \left. \frac{1}{2}m_{2} \right| 
    1\lambda\frac{}{} \right\rangle & \leftrightarrow & 
    \eta_{2}^{\dagger}
    \left( \frac{-1}{\sqrt{2}}\,
    \bbox{\sigma}\cdot\bbox{\epsilon}^{(\lambda)}\right)\eta_{1c} \\
    \left\langle \frac{1}{2}m_{2} \left|\right. 1\lambda\, \frac{1}{2}m_{1} 
    \right\rangle & \leftrightarrow & 
    \eta_{2}^{\dagger}
    \left( \frac{1}{\sqrt{3}}\,
    \bbox{\sigma}\cdot\bbox{\epsilon}^{(\lambda)}\right)\eta_{1} \\
    \left\langle 00 \right|\left. 1\mu_{1}1\mu_{2} \right\rangle 
    & \leftrightarrow &
    \frac{-1}{\sqrt{3}}\,
    \bbox{\epsilon}^{(\mu_{1})}\cdot\bbox{\epsilon}^{(\mu_{2})} \\
    \left\langle 1\lambda \right|\left. 1\mu_{1}1\mu_{2} \right\rangle 
    & \leftrightarrow & 
    \bbox{\epsilon}^{(\lambda)\dagger}\cdot\left( 
    \frac{-i}{\sqrt{2}}\,
    \bbox{\epsilon}^{(\mu_{1})}\bbox{\times\epsilon}^{(\mu_{2})} \right)
\end{eqnarray}

\subsection{Relation to the SAID formalism}
The $\bbox{\sigma}$-$\bbox{\epsilon}$ formalism uses the normalization
\begin{equation}
    \frac{{\rm d}\bar{\sigma}}{{\rm d}\Omega}(\pi^{+}d\rightarrow pp) =
    \frac{m^{2}}{4(2\pi)^{2}s_{\pi{d}}}\frac{p}{k}\,\frac{1}{3}
    \sum_{{\rm spins}} |\mathcal{M}|^{2},
\end{equation}
where $\sqrt{s_{\pi{d}}}$ is the cm energy. 
The matrix element is then of the form
\begin{equation}
    \mathcal{M} = \left<pp\left|\sum_i\mathcal{A}_{i}
    \mathcal{O}_i\right|\pi^+d\right>\:,
\end{equation}
where $\mathcal{A}_i$ is the complex amplitude for a particular partial wave. 

In the formalism of the SAID database~\cite{SAID}
\begin{equation}
    \frac{{\rm d}\bar{\sigma}}{{\rm d}\Omega}(\pi^{+}d\rightarrow pp) =
    \frac{1}{6k^{2}}\sum_{i=1}^{6} |H_{i}|^{2},
\end{equation}
where the helicity amplitudes, $H_{i}$, are linear combinations of partial wave
amplitudes $T$. The relation between the $\bbox{\sigma}$-$\bbox{\epsilon}$ and 
SAID partial wave amplitudes is then
\begin{equation}
    \mathcal{A} = \frac{2\pi}{m}\sqrt{\frac{2s_{\pi{d}}}{pk}} T,
\end{equation}
since all the spin-momenta structures and relative phases are incorporated in
the $\bbox{\sigma}$-$\bbox{\epsilon}$ formulae.


\begin{figure}[p]
\centering    \includegraphics{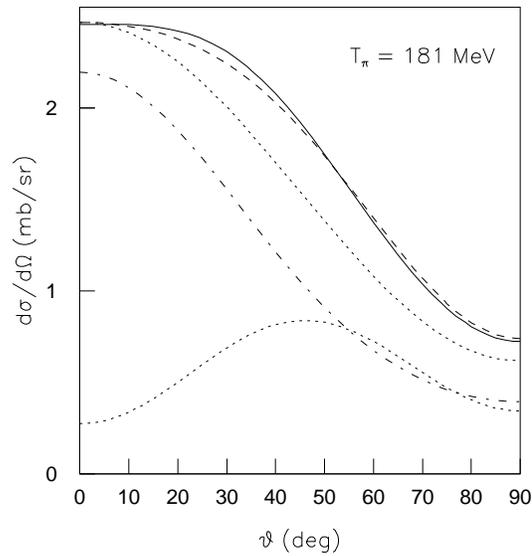}
\caption{Differential cross section for the $\pi^{+}d\rightarrow{pp}$ reaction 
at \mbox{$T_{\pi}$ = 181 MeV}, calculated with nine amplitudes (dashed line) 
compared to the differential cross section obtained from the SAID database 
(solid line). The cross sections for $^{1}\!D_{2}p$ (uppermost dotted line) and
the singlet (dot-dashed line) and triplet (lowest dotted line) amplitudes are 
also given.}
\label{fig:ppdpi}
\end{figure}

\begin{table}[p]
\centering
\begin{tabular}{ccccccc}
    $T_{d}$ (MeV) & 787 & 940 & 1084 & 1250 & 1412 & 1938 \\ 
    \hline
    $\sigma_{dd}$ (${\rm mb}$) & 
    118 & 122 & 128 & 137 & 145 & 152 \\
    $\frac{1}{2}(\sigma_{pn}+\sigma_{pp})$ (${\rm mb}$) 
    & 28 & 33 & 36 & 39 & 42 & 44 \\
    $\sigma_{\pi\alpha}$ (${\rm mb}$) & 100 & 235 & 328 & 332 & 275 & 135 \\
    $\sigma_{\pi{d}}$ (${\rm mb}$) & 
    42 & 96 & 176 & 229 & 185 & 71 \\
    $\mathcal{D}({\rm estimate})$ & 1.50 & 1.38 & 0.77 & 0.46 & 0.48 & 0.78 \\
    $\mathcal{D}({\rm fit})$ & 2.2 & 0.90 & 0.72 & 0.73 & 0.80 & 1.0 \\
\end{tabular}

\caption{Distortion factor $\mathcal{D}$ calculated according to 
eq.~(\ref{eq:dist}) for different $dd$ energies. The total cross section data 
are from Refs~\protect\cite{PDG,Kishida,Pedroni,pialpha}. The estimated 
reduction factor is compared to the scale factor needed to fit our predictions 
to data.}
\label{tab:dist}
\end{table}

\end{document}